# Attributes of GRB Pulses: Bayesian Blocks Analysis of TTE Data; a Microburst in GRB 920229

Jeffrey D. Scargle,[1] Jay Norris,[2] and Jerry Bonnell[3]

[1] *NASA/Ames Research Center, Moffett Field, CA 94035*
[2] *NASA/Goddard Space Flight Center, Greenbelt, MD 20771*
[3] *Universities Space Research Association*

**Abstract. Bayesian Blocks** is a new time series algorithm for detecting localized structures (*spikes* or *shots*), revealing pulse shapes, and generally characterizing intensity variations. It maps raw counting data into a maximum likelihood piecewise constant representation of the underlying signal. This bin-free method imposes no lower limit on measurable time scales. Applied to BATSE TTE data, it reveals the shortest know burst structure – a spike superimposed on the main burst in GRB 920229 = Trigger 1453, with rise and decay timescales $\approx$ few100$\mu$sec.

## GRB TIME PROFILES

While there is phenomenological understanding of some general features of the incredibly varied GRB time profiles [10,16,6,1,3,17], the underlying physical processes are unknown. We are analyzing BATSE data with a variety of techniques aimed at extracting physically relevant information without regard to physical models.

The BATSE time-tagged event, or TTE, data have the highest time resolution (but are perforce limited to short bursts). We have developed a bin-free time domain method, based on a rigorous statistical treatment of the Poisson photon detection process, to characterize the underlying signal in spite of the large statistical fluctuations inherent in photon counting. Called Bayesian Blocks (**BB**) on account of the nature of its statistical foundation, the technique starts from raw photon counts and produces the most likely piecewise constant representation of brightness as a function of time. In this paper we treat only TTE data, but elsewhere the same procedure is developed for binned and TTS data [12].

## THE ANALYSIS METHOD: BAYESIAN BLOCKS

The idea rests on the simplicity of mathematical models of two processes: a source of constant brightness over a given time interval (this model is here denoted $\mathcal{M}_1$), and the detection of an individual photon. These simplicities lead to an exact formula for the model likelihood for a given array of photon arrival times. Marginalization of nuisance parameters with physically derived priors [12,13] yields the *global likelihood*

$$\mathcal{L}(\mathcal{M}_1|D_{TTE}) = \phi(N, M) = \frac{\Gamma(N+1)\Gamma(M-N+1)}{\Gamma(M+2)} \tag{1}$$

where $N$ is the number of photons detected in the interval, $M$ is the length of the interval in the units in which arrival times are quantized, and $\Gamma(N+1) = N!$ is the gamma function. When compared to the analogous global likelihood of a second model, this quantity measures the probability of the constant source model.

Indeed, the other model we are interested in, $\mathcal{M}_2$, is that in which the observed interval is divided into two subintervals, separated by a *change point*, each of which has a constant Poisson rate. The global likelihood of this compound model is

$$\mathcal{L}(\mathcal{M}_2|D_{TTE}) = \sum_{n_{cp}} \phi(N_1, M_1)\phi(N_2, M_2) \tag{2}$$

where the pairs $N, M$ are the number of photons in, and length of, each subinterval, and the sum over $n_{cp}$ marginalizes the change point location.

Taking $\rho$ to be the ratio of the prior probabilities of segmented and unsegmented models, the odds ratio

$$O = \rho \frac{\mathcal{L}(\mathcal{M}_2|D_{TTE})}{\mathcal{L}(\mathcal{M}_1|D_{TTE})} \tag{3}$$

gives the relative posterior probabilities of the two models (for the same data). The choice $\rho = 1$ expresses lack of prior knowledge of the structure. If the odds ratio in (3) favors segmentation, the location of the optimal change point is found by maximizing the summand in the eq. (2). This is easy, because the summand can be written as a function of only one variable – the location of the photon at the change point. In turn the maximum likelihood Poisson rates are found to be $\frac{N}{M}$ for each subinterval. The full procedure to form a block representation of the signal consists of iterative application of the decision making step just described, until each subinterval is judged better off left undivided.

We are using this algorithm [11] to detect and characterize pulse structures known to make up the time-profiles of many $\gamma$-ray bursts [10]. The **BB** decomposition has proved to be an excellent, completely automatic, way to determine approximate widths, locations, and amplitudes of pulses, without invoking parametric or other explicit pulse-shape models. These are fed to a nonlinear optimizer which deconvolves, and estimates parametric models of, overlapping pulses. Moreover the information contained in the blocks is itself useful, especially in cases where the overlap is not too severe.

## A MICROBURST IN GRB 920229 = TRIGGER 1453

Trigger 1453 nicely displays the time resolution and dynamic range of the **BB** procedure. From the time of the First BATSE Catalog [4] it has been commonly known that this burst contains a spike of duration $<\approx 10$ ms near the end of a relatively flat burst of duration $\approx 0.192$ sec.

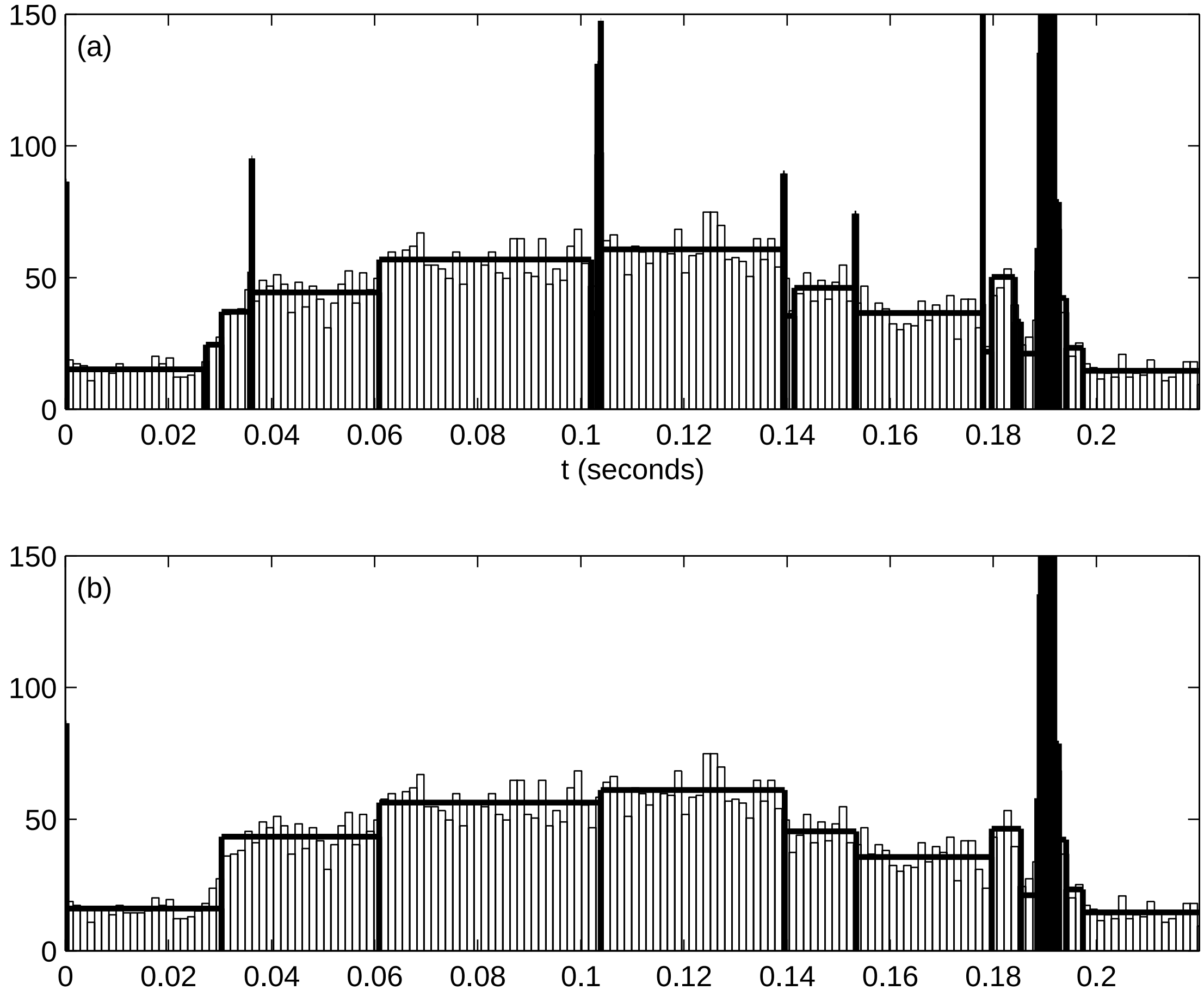

**FIGURE 1.** Bayesian Block decomposition for Trigger 1453, all energy channels. Count rate (in units of 1000 c/sec) *vs* time (sec). BB: thick lines; binned TTE data (for display only): thin lines. The microburst peak at $0.191s$ sec is offscale. The last BB shown extends to the end of the data at 1.42 sec, and is a good estimate of the background. (a) $\rho = 1$; (b) $\rho = .01$.

With no special treatment this spike showed up as the shortest in the distribution of block widths. Figure 1 shows the **BB** decomposition of the TTE data. The the top panel, for $\rho = 1$, shows tall thin blocks at several places on the broad plateau. These undoubtedly spurious *whiskers* are probably signs of occasional errors in the segmentation decision, based on $O > 1$ *vs* $O <= 1$ in eq. 3; for $O \approx 1$ this coin flip should go wrong about half the time. Taking $\rho < 1$ wards off these errors, as seen in the bottom panel; real structures, including the (offscale) spike, typically correspond to astronomically large odds ratios and are therefore barely affected.

Figure 2 includes individual energy channels and expands the region of the spike – confirming its reality, except in the low S/N channel 4. This is the only short spike in our BB analysis of 317 short bursts, in agreement with a search for spikes in the range $10\mu s - 10ms$ in TTE data for 20 bright bursts [15].

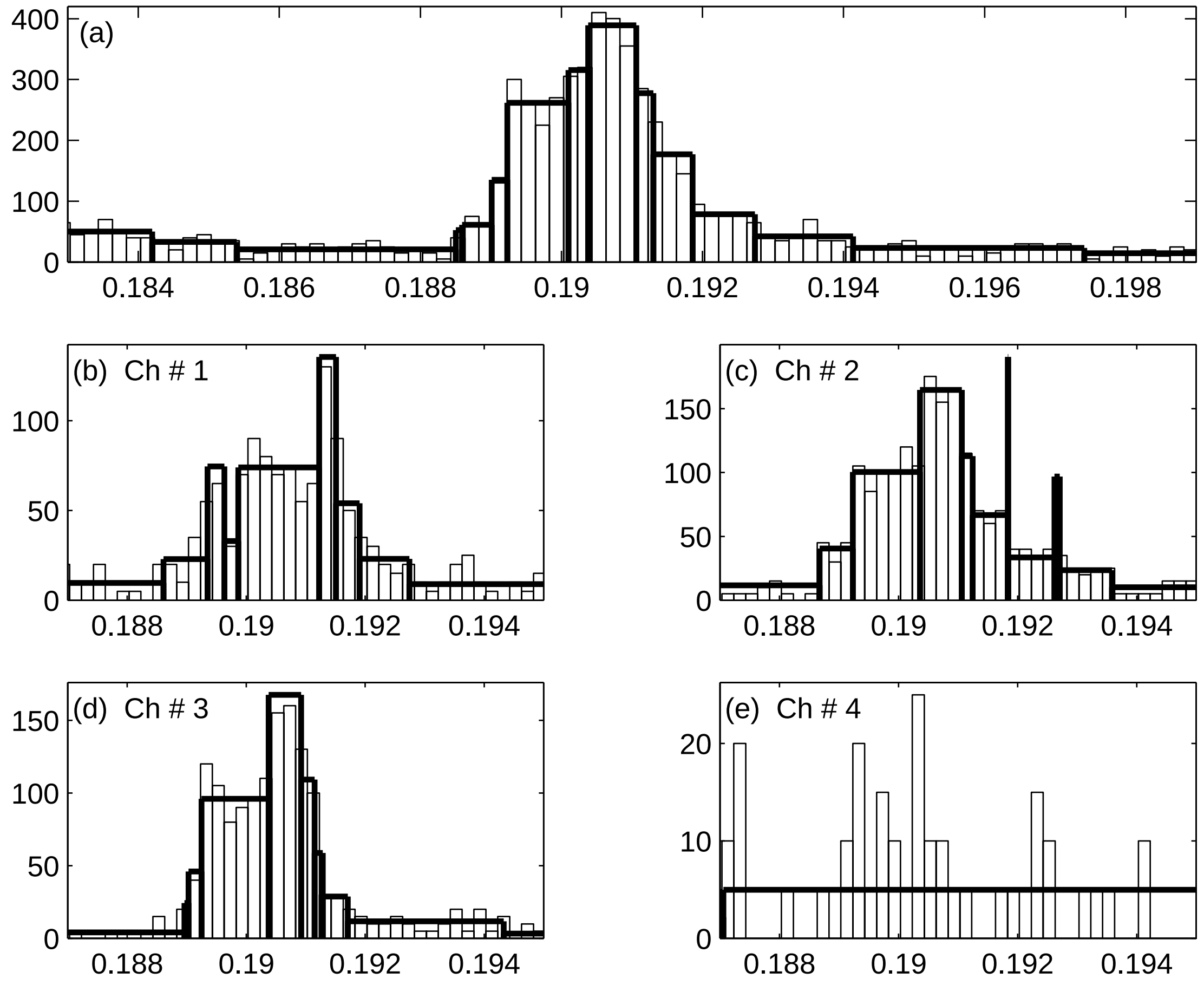

**FIGURE 2.** The microburst as in Fig 1; $\rho = 1$. (a) sum of energy channels; (b-d) channels 1-4.

## OTHER WORK

A more detailed study of this burst will address – using a sensitive Bayesian method [18] for assessing the likelihood that data histograms are from the same distribution – whether the detector (or energy) distributions of spike photons are different from those of the main burst. This seems unlikely, but would argue that the spike, in a different part of the sky, is different from ordinary bursts.

We are investigating replacement of the *ad hoc* iteration used here with the more rigorous approach of multiple change point detection [2]. While this may lead to improvements, numerical experiments with real and synthetic data show our implementation of **BB** to be accurate and efficient. Other relevant work includes a negative search for microsecond flares [14], a study of fundamental instrumental limits on detection of short spikes [9], a Bayesian method to detect bursts in binned count data [7], and wavelet methods [5].

Thanks to Tom Loredo, for many suggestions, and NASA's Astrophysics Data Program.

## REFERENCES

1. Brock, M., *et al*, in Proc. 2nd GRB Workshop, eds. G. Fishman, J. Brainerd, and K. Hurley (New York: AIP) **307**, (1994).
2. Chib, S., "Estimation and Comparison of Multiple Change Point Models," J. Econometrics, in press (1996).
3. Fenimore, E., Madras, C., and Nayakshin, S., ApJ **473**, 998 (1996).
4. Fishman *et al.*, Apj Supp **92**, 229 (1994).
5. Kolaczyk, E., "Nonparametric Estimation of Gamma-Ray Burst Intensities Using Haar Wavelets," ApJ **483**, 34 (1997); references in [12].
6. Li, H., and Fenimore, E., ApJ Lett **469**, L115 (1996).
7. Marsden, D., and Rothschild, R., "Detection of Bursts in Time Series Data Using Bayesian Techniques," AAS/HEAD Meeting, session Bayes. Stat. in Ap (1997).
8. Mitrofanov, I., AP & SS **155**, 141 (1989)
9. Nemiroff, R., Norris, J., Bonnell, J., and Marani, G., "GRB Spikes Could Resolve Stars," submitted to ApJ (1997).
10. Norris, J., Nemiroff, R., Bonnell, J., Scargle, J., Kouveliotou, C., Paciesas, W., Meegan, C., and Fishman, G., ApJ **459**, 393 (1996).
11. Scargle, J., Norris, J, and Bonnell, J., "Attributes of Gamma-Ray Burst Pulses: I. Short Bursts Analyzed with BATSE TTE Data," in preparation.
12. Scargle, J., "Studies in Astronomical Time Series Analysis: V. Bayesian Blocks, A New Method to Analyze Structure in Photon Counting Data" submitted to ApJ (1997); `http://xxx.lanl.gov/abs/astro-ph/9711233`.
13. Scargle, J., and Bloom, E., "Bayesian Blocks, A New Method to Analyze Structure in Photon Counting Data" AAS/HEAD Meeting, session Bayes. Stat. in Ap (1997).
14. Schaefer, B.,*et al.*., Ap J **404** 673 (1993).
15. Schaefer, B., Walker, C., and Leyder, A., "A Flare with a 200 Microsecond Risetime, but No Microsecond Blackbody Emission," this conference (1997).
16. Stern, B., ApJ Lett **464**, L111 (1996).
17. Stern, B., and Svensson, R., ApJ Lett **469**, L109 (1996).
18. Wolpert, D., "Determining Whether Two Data Sets are from the same Distribution," 1996, 271-276, in *Maximum Entropy and Bayesian Methods*, Hanson, K. and Silver, R., eds., Kluwer; and Loredo, unpublished.